\documentclass[a4paper,11pt]{article}
\usepackage{pos}

\usepackage{hyperref}
\usepackage{bbold}
\usepackage{amsmath}
\usepackage{multirow}
\usepackage{graphicx}
\usepackage{xfrac}
\usepackage[utf8]{inputenc}

\usepackage{extarrows}

\newcommand{\be}{\begin{equation}}
\newcommand{\ee}{\end{equation}}
\newcommand{\bea}{\begin{eqnarray}}
\newcommand{\eea}{\end{eqnarray}}

\newcommand{\Dlr}{\buildrel \leftrightarrow \over D\raise-1pt\hbox{}}

\newcommand{\avgxN}{\langle x^N \rangle}
\newcommand{\avgx}{\langle x \rangle}
\newcommand{\avgxx}{\langle x^2 \rangle}
\newcommand{\avgxxx}{\langle x^3 \rangle}
\newcommand{\avgxxxx}{\langle x^4 \rangle}
\newcommand{\avgxxxxx}{\langle x^5 \rangle}
\newcommand{\avgxxxxxx}{\langle x^6 \rangle}

\begin{document}
\title{$x$-dependence reconstruction of pion and kaon PDFs from Mellin moments
author}
\ShortTitle{Pion and kaon PDF reconstruction from Mellin moments}
\author[a,b]{Constantia Alexandrou}
\author[b]{Simone Bacchio}
\author[c]{Ian Clo\"et}
\author[d]{Martha Constantinou}
\author[a,b]{Kyriakos Hadjiyiannakou}
\author[b]{Giannis Koutsou}
\author*[c]{Colin Lauer}
\affiliation[a]{Department of Physics, University of Cyprus,  P.O. Box 20537,  1678 Nicosia, Cyprus}
\affiliation[b]{Computation-based Science and Technology Research Center,
  The Cyprus Institute, 20 Kavafi Str., Nicosia 2121, Cyprus}
\affiliation[c]{Physics Division, Argonne National Laboratory, Lemont, Illinois 60439, USA}
\affiliation[d]{Department of Physics, Temple University, 1925 N. 12th Street, Philadelphia, PA 19122-1801, USA}
\emailAdd{colin.j.lauer@temple.edu}
\abstract{We present a calculation of the connected-diagram contributions to the
first three non-trivial Mellin moments for the pion and kaon, extracted using local operators with up to 3 covariant
derivatives. We use one ensemble of gauge configurations with two
degenerate light, a strange and a charm quark ($N_f$=2+1+1) of
maximally twisted mass fermions with clover improvement. The ensemble
has a pion mass $\sim$260 MeV, and a kaon mass $\sim$530 MeV. We reconstruct the $x$-dependence of the
PDFs via fits to our results, and find that our lattice data favor a $(1-x)^2$-behavior in the large-$x$ region for both the pion and kaon PDFs. We integrate the
reconstructed PDFs to extract the higher moments, $\langle x^n \rangle$, with $4 \leq n \leq 6$.
Finally, we compare the pion and kaon PDFs, as well as the ratios of
their Mellin moments, to address the effect of SU(3) flavor symmetry
breaking.}
\FullConference{The 38th International Symposium on Lattice Field Theory\\
Virtual\\
July 26-30, 2021
}

\maketitle

\section{Introduction}
Pion, kaons, and their structures are interesting topics of study in particle physics. Since pions and kaons are both pseudoscalar particles and have similar valence quark structures---the kaon has one light and one strange valence quark instead of two light quarks---comparing their structures can give insights into relations between strong interactions and quark mass effects~\cite{Hutauruk:2016sug}, as well as SU(3) flavor symmetry breaking effects which are caused by the heavier strange quark compared to light quarks~\cite{Zyla:2020zbs}. There is some existing experimental data from pion induced Drell-Yan~\cite{Conway:1989fs}, but kaon data is more limited and cannot make definite conclusions about SU(3) symmetry breaking effects. The $x$-dependence of PDFs has been calculated using Lattice Quantum Chromodynamics (LQCD) (see, e.g. the reviews of Refs.~\cite{Cichy:2018mum,Constantinou:2020pek,Cichy:2021lih}), but pions and kaons have not been as studied extensively as the nucleons.

In these proceedings, we use Mellin moments of PDFs to reconstruct the full $x$-dependence of the pion and kaon PDFs. Reconstructing PDFs from their Mellin moments calculated on the lattice has often been argued to be challenging and even unfeasible. Higher moments include higher uncertainties due to derivatives in the local operators, which introduce more gauge noise, as well as increased noise due to the need for nonzero momentum frames. Additionally, mixing among lower dimensional operators under renormalization is unavoidable when calculating moments higher than $\avgxxx$. The combination of these challenges has caused early attempts at reconstructing PDFs~\cite{Detmold:2001dv,Holt_2010} to be unable to conclusively determine its feasibility. 

Even though these challenges exist, there have been advancements made to computers, algorithms, and computational methods so that Mellin moments of PDFs can be calculated more precisely. In Refs.~\cite{Alexandrou:2020gxs,Alexandrou:2021}, we calculated the first three nontrivial Mellin moments of the pion and kaon PDFs with the excited state effects reliably removed and without mixing with lower dimensional operators. These improvements  motivate us to attempt to reconstruct the $x$-dependence of the PDFs with three goals: \textbf{a.} test the feasibility and limitations reconstructing the PDFs, \textbf{b.} observe the behavior of the reconstructed PDFs at large-$x$, and \textbf{c.} use the reconstructed PDFs to calculate the higher moments with $n>3$. 

\section{Theoretical and Lattice Setup\label{sec:II}}

For our calculations we use a gauge ensemble produced by the Extended Twisted Mass Collaboration (ETMC)~\cite{Alexandrou:2018egz}. The ensemble, labeled cA211.30.32, uses $N_f=2+1+1$ twisted clover fermions with clover improvement and the Iwasaki improved gluon action. Table~\ref{tab:params} contains the key ensemble parameters. The statistics used in the Mellin moment calculations are given in Table~\ref{tab:statistics}.

{\small{
\begin{table}[h!]
\centering
\renewcommand{\arraystretch}{1.2}
\renewcommand{\tabcolsep}{6pt}
\begin{tabular}{| l| c | c | c | c | c  | c | c |}
    \hline
    \multicolumn{8}{|c|}{Parameters} \\
    \hline
    Ensemble   & $\beta$ & $a$ [fm] & volume $L^3\times T$ & $N_f$ & $m_\pi$ [MeV] &
    $L m_\pi$ & $L$ [fm]\\
    \hline
    cA211.30.32 & 1.726 & 0.093  & $32^3\times 64$  & $u, d, s, c$ & 260
    & 4 & 3.0 \\
    \hline
\end{tabular}
\caption{Parameters of the ensemble used in this study.}
\label{tab:params}
\end{table}
}}

{\small{
\begin{table}[h!]
\centering
\renewcommand{\arraystretch}{1.2}
\renewcommand{\tabcolsep}{6pt}
\begin{tabular}{| c | c | c | c | c |}
    \hline
    \multicolumn{5}{|c|}{Statistics} \\
    \hline
    $t_s/a$ & $\#$ configurations   & $\#$ source positions & $\#$ momentum boost & Total \\
    \hline
    \multicolumn{5}{|c|}{$\avgx$ two-point correlators} \\
    \hline
    --- & 122 & 16 & 1 & 1,952 \\
    \hline
    \multicolumn{5}{|c|}{$\avgx$ three-point correlators} \\
    \hline
    12, 14, 16, 18, 20, 24 & 122 & 16 & 1 & 1,952 \\
    \hline
    \multicolumn{5}{|c|}{$\avgxx$ $\&$ $\avgxxx$ two-point correlators} \\
    \hline
    --- & 122 & 72 & 8 & 70,272 \\
    \hline
    \multicolumn{5}{|c|}{$\avgxx$ $\&$ $\avgxxx$ three-point correlators} \\
    \hline
    12 & 122 & 16 & 8 & 15,616 \\
    14, 16, 18 & 122 & 72 & 8 & 70,272 \\
    \hline
\end{tabular}
\caption{Statistics used in the Mellin moment calculations.}
\label{tab:statistics}
\vspace*{0.2cm}
\end{table}
}}
We calculate the moments at various values of the source-sink separation $t_s$ and perform a two-state fit in order to control excited-state contamination. Refs.~\cite{Alexandrou:2020gxs,Alexandrou:2021} contain a full description of the process we use to calculate the PDF moments.
In our reconstruction of the pion and kaon PDFs, we use the first three non-trivial Mellin moments using the two-state fit results, which we summarize in Table~\ref{tab:moments}. 

{\small{
\begin{table}[h!]
\begin{center}
    \begin{tabular}{cccc}
        \hline\hline\\[-2.5ex]
        method & $\,\,\avgx_{u^+}^{\pi}$ & $\,\,\avgx_u^{k}$ & $\,\,\avgx_s^{k}$ \\[0.5ex]
        \hline\\[-2.5ex]
        2-state     &$\,\,\,$0.261(3)   &$\,\,\,$0.246(2)   &$\,\,\,$0.317(2)   \\\hline\\
    \end{tabular}  
    \begin{tabular}{ccccccc}
        \hline
        method & $\,\,\avgxx_{u^+}^{\pi}$ & $\,\,\avgxx_u^{k}$ & $\,\,\avgxx_s^{k}$ & $\,\,\avgxxx_{u^+}^{\pi}$ & $\,\,\avgxxx_u^{k}$ & $\,\,\avgxxx_s^{k}$ \\[0.5ex]
        \hline\\[-2.5ex]
        2-state     &$\,\,\,$0.110(7)  &$\,\,\,$0.096(2)   &$\,\,\,$0.139(2)  &$\,\,\,$0.024(18) &$\,\,\,$0.033(6)   &$\,\,\,$0.073(5)      \\ [0.5ex]
        \hline
    \end{tabular}  
    \caption{Final results for the first three non-trivial pion and kaon Mellin moments obtained from the one- and two-state fits.}
    \label{tab:moments}
\end{center}
\end{table}
}}

\section{Reconstruction of PDFs}

We start the reconstruction of the $x$-dependent PDFs, $q^f_M(x)$, with the standard functional form
\begin{equation}
\label{eq:PDF}
q^f_M(x) = N x^\alpha (1-x)^\beta (1+\rho \sqrt{x} + \gamma x)   \,, 
\end{equation} 
 where $(M,f)=(\pi,\,u),\,(K,\,u),\,(K,\,s)$. $N$ is a normalization defined by charge conservation,
which leads to
\begin{equation}
\label{eq:Norm}
N= \frac{1}{B(\alpha+1,\beta+1) + \gamma B(2+\alpha,\beta+1) }\,,
\end{equation} 
where $B$ is the Euler beta-function. The fit parameters in Eq.~(\ref{eq:PDF}) are $\alpha$, $\beta$, $\gamma$ and $\rho$. We set the parameter $\rho$ to be zero, since it is generally assumed to be negligible~\cite{Chen:2016sno}. The $n^{\rm th}$-moment is calculated by taking the integral of Eq.~(\ref{eq:PDF}), i.e.,
\begin{equation}
\langle x^n \rangle = N \int_0^1 dx\, x^\alpha (1-x)^\beta (1 + \gamma x) \,,
\end{equation} 
which results in
\begin{equation}
\label{eq:moments}
\displaystyle
\langle x^n \rangle = \frac{ \Big{(}\prod_{i=1}^n (i+\alpha) \Big{)} \, \Big{(}n+2+\alpha+\beta+(i+1+\alpha) \gamma\Big{)}}
{\Big{(}\prod_{i=1}^n (i+2+\alpha+\beta) \Big{)}\, \Big{(}2+\alpha+\beta+(1+\alpha)\gamma\Big{)}} \,,\quad n>0\,.
\end{equation}
We calculate the fit parameters in Eq.~(\ref{eq:moments}), using our Mellin moment results given in Table~\ref{tab:moments} as inputs. We use NNLO expressions to evolve our Mellin moment results to a scale of 5.2 GeV so that comparison can be made to results from global fits and models.

We first test how the reconstructed PDF depends on the number of fit parameters by performing a 2-parameter ($\gamma=0$) and a 3-parameter fit. We find that the reconstructed PDF are compatible for the two kind of fits, and choose the 2-parameter fit as final method. The obtained fit parameters are given in Table~\ref{table:fit_param_pion_u}. The fits for the pion are less stable than those for the kaon in both cases because the gauge noise is larger in the pion. 

\begin{table}[h!]
    \centering
    \begin{tabular}{ccc}
      \hline\hline\\[-2.5ex]
type & $\qquad\alpha\qquad$ & $\qquad\beta\qquad$  \\[0.5ex]
\hline\\[-2.5ex]
$\pi^u$  &$\,\,\,$-0.05(19)   &$\,\,\,$2.20(64)    \\ [0.5ex]
$K^u$    &$\,\,\,$-0.005(81)   &$\,\,\,$ 2.59(28)   \\ [0.5ex]
$K^s$    &$\,\,\,$0.26(9)   &$\,\,\,$2.27(22)    \\ [0.5ex]
      \hline\hline
    \end{tabular}
        \caption{The fit parameters for $q_\pi^u$, $q_K^u$ and  $q_K^s$ at 5.2 GeV. The error in the parenthesis is statistical.}
        \label{table:fit_param_pion_u}
  \end{table}

\begin{figure}[!h]
    \centering
    \includegraphics[width=\linewidth]{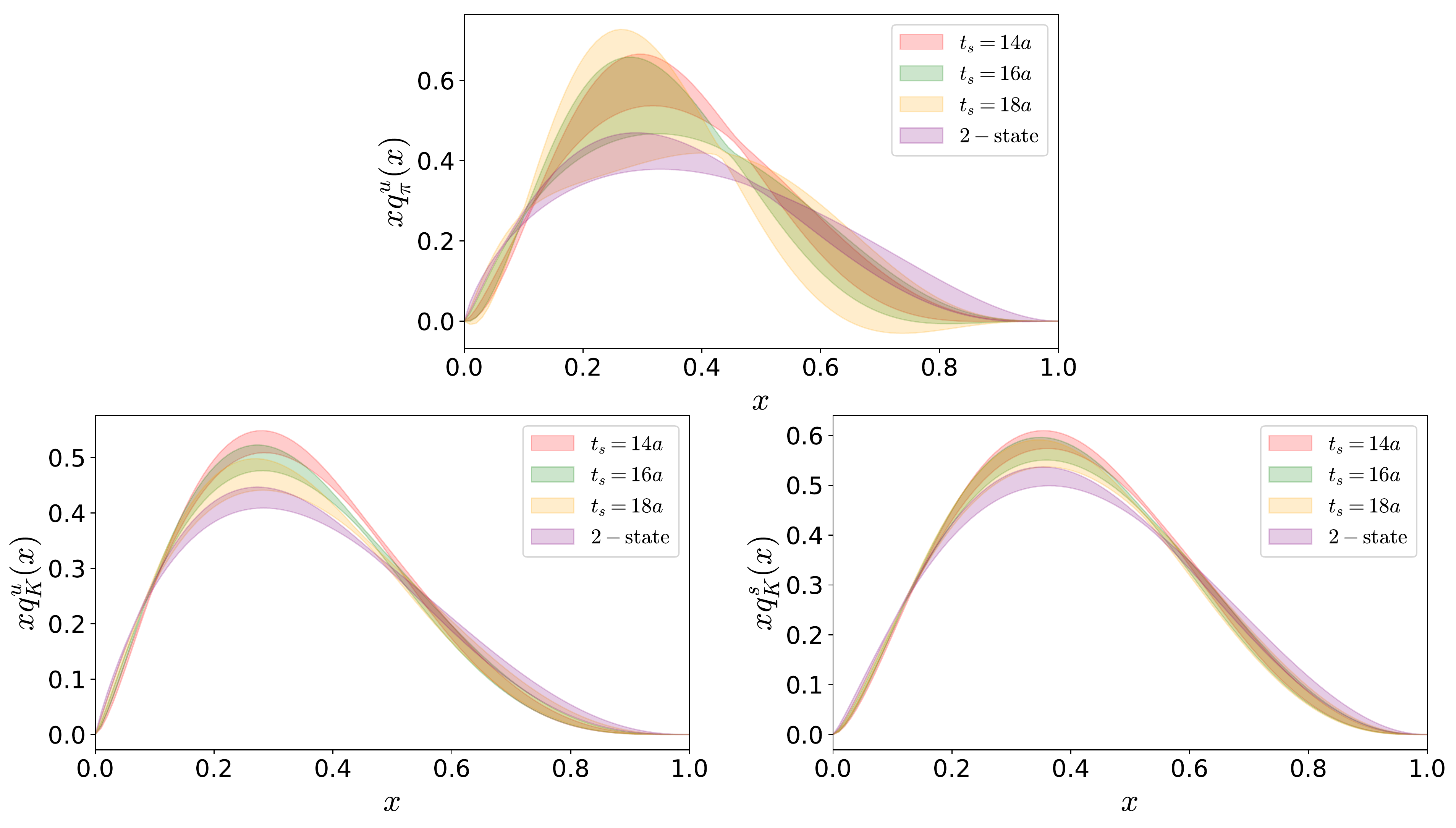}
    \vspace*{-0.2cm}
    \caption{The $x$ dependence of $x q_\pi^u(x)$ (top), $x q_K^u(x)$ (lower left panel), and $x q_K^s(x)$ (lower right panel) with $t_s/a=$14, 16, 18, and the 2-state fit shown as the pink, green, yellow and purple bands, respectively. The results are given in the $\overline{\rm MS}$ scheme at a scale of 27 GeV$^2$.}
    \label{fig:xPDF_tsink}
\end{figure}

Using Eq.~(\ref{eq:PDF}) we reconstruct the $x$-dependent PDFs and study the effects of excited-states contamination on $q_M^f(x)$. Such effects may be non-trivial due to the moments having a non-linear dependence on the fit parameters in Eq.~(\ref{eq:PDF}). We perform a 2-parameter fit on our Mellin moment results at $t_s/a=14 - 18$, as well as the two-state fit and reconstruct the PDFs, as shown in Fig.~\ref{fig:xPDF_tsink}.  We find that the PDFs increase with lower $t_s$, meaning that the excited-state contamination leads to a higher PDF. We see a nice convergence as $t_s$ increases with the two-state fit values eventually converging with a peak around $x\sim 0.3 - 0.4$, i.e., $x q_\pi^u(0.3)\sim 0.4$, $x q_K^u(0.3)\sim 0.4$ and $x q_K^s(0.4)\sim 0.5$. We use the PDFs extracted form the two-state fit results for the moments (purple band in Fig.~\ref{fig:xPDF_tsink}) as our final results. 

\begin{figure}[!h]
    \centering
    \includegraphics[scale=0.4]{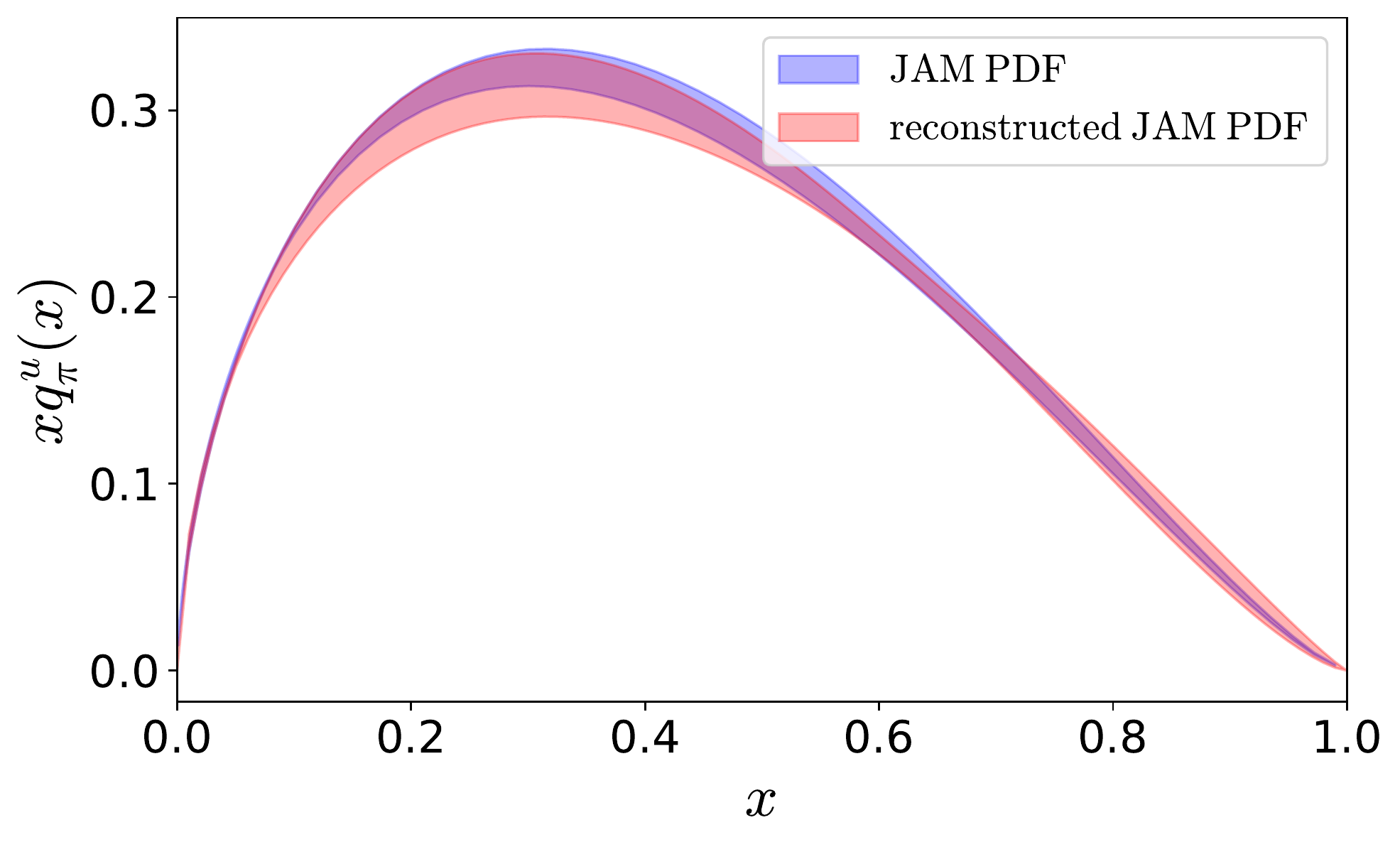}
    \vspace*{-0.5cm}
    \caption{Pion PDF as calculated by the JAM collaboration (blue band) and the PDF reconstructed from its moments with $n \leq 3$ (pink band). The reported scale is 27 GeV$^2$.}
    \label{fig:PDF_JAM}
\end{figure}

Since we reconstruct the PDFs using only the $n \leq 3$ moments, we test whether three moments are enough to successfully calculate the $x$-dependence. To do this, we use the PDF and its moments calculated by the Jefferson Lab Angular Momentum (JAM) collaboration ~\cite{Barry:2018ort}. We use the same procedure as we did previously with our lattice data to reconstruct the JAM PDF, now using as inputs the JAM moments with $n \leq 3$. In Fig.~\ref{fig:PDF_JAM}, the reconstructed PDF and original JAM PDFs are compared to each other. We find that the two agree well for almost all regions of $x$ within uncertainties. The reconstructed PDF has much larger uncertainties, due to the lost information from truncating the moments at $n=3$. As another check, we calculate the $n=4$ moment using the fitting parameters and Eq.~(\ref{eq:moments}). We find $\avgxxxx_\pi^u=0.026(2)$, which is in excellent agreement with the moment calculated using the JAM framework, $\avgxxxx_\pi^u=0.027(2)$ supporting the fact that the PDFs can feasibly be reconstructed using the $n \leq 3$ moments. 

\begin{figure}[!h]
    \centering
    \includegraphics[width=\linewidth]{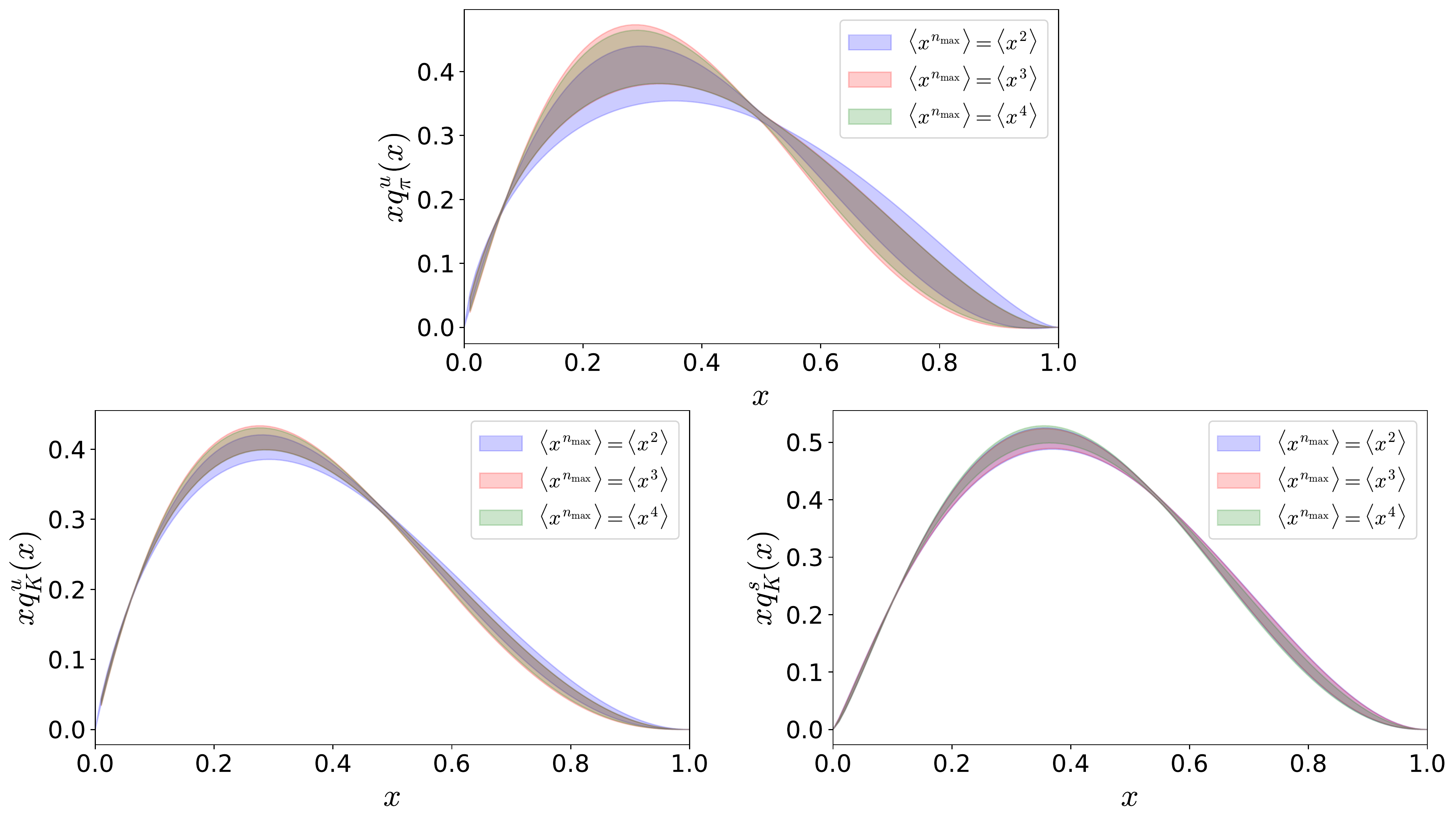}
    \caption{Top: The $x$ dependence of $x q_\pi^u(x)$ using the 2-parameter fit with $\langle x^n_\mathrm{max} \rangle=\avgxx,\,\avgxxx,\,\avgxxxx$ The blue, pink and green bands represent $n_\mathrm{max}=2$, 3, and 4, respectively. For the green band, we use $\avgxxxx_\pi^u$~\cite{Barry:2018ort} as a constraint. Bottom: Same as top panel for $x q_K^u(x)$ (left panel) and $x q_K^s(x)$ (right panel) using the BLFQ-NJL~\cite{Lan:2019rba} $\avgxxxx_K$ values as constraints for the green band.}
    \label{fig:PDF_x4_constraint}
\end{figure}

We further test the feasibility of reconstructing the PDFs by performing fits including moments up to $\langle x^{n_\textrm{max}} \rangle$, with $n_\textrm{max}=2$, 3, or 4. We only use lattice data for $n_\textrm{max}=2$ and 3, and add another constraint for $n_\textrm{max}=4$ by using the value of $\avgxxxx$ obtained from global fits and models for the pion and kaon, respectively. For the pion, we use $\avgxxxx_\pi^u=0.027(2)$ from the JAM analysis~\cite{Barry:2018ort}, and $\avgxxxx_K^s=0.029^{+0.005}_{-0.004}$, $\avgxxxx_K^u=0.021^{+0.003}_{-0.003}$ from BLFQ-NJL~\cite{Lan:2019rba} for the kaon. We compare the PDFs reconstructed with $n_\textrm{max}=2$, 3, and 4 in Fig.~\ref{fig:PDF_x4_constraint}. The constraint of the PDFs is improved with the addition of $n=3$ and, interestingly, the shape of the PDFs are not affected by the addition of $n=4$. Therefore, the effect of including higher moments in the fits is within the shown uncertainties and we use as our final estimates $n_\mathrm{max}=3$. To summarize our final estimates, we choose the PDFs reconstructed from a 2-parameter fit with our two-state fit lattice results up to $\avgxxx$ as inputs.

\begin{figure}[!h]
    \centering
    \includegraphics[scale=0.4]{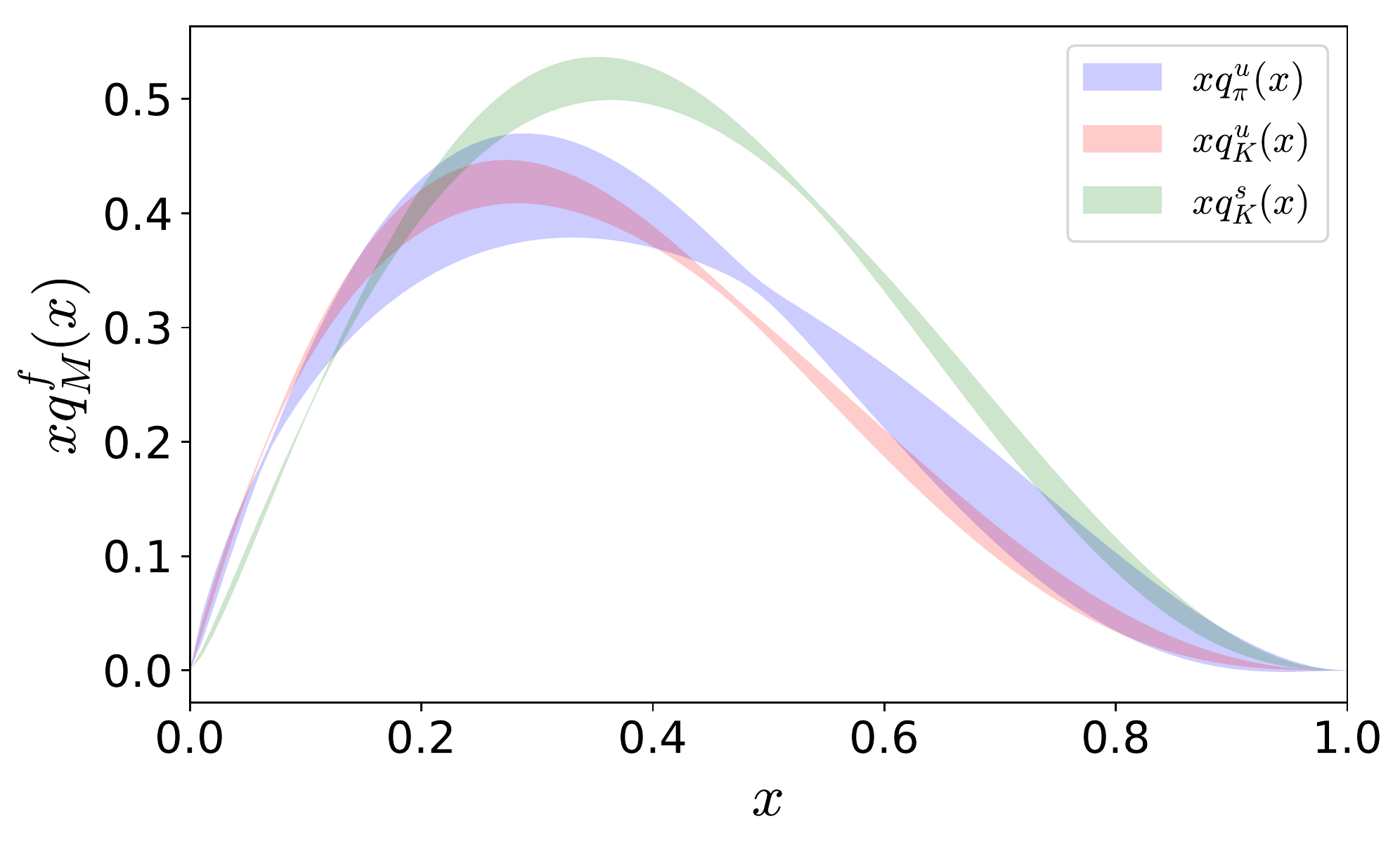}
    \caption{Comparison of $x q_\pi^u(x)$ (blue band), $x q_K^u(x)$ (pink band) and $x q_K^s(x)$ (green band) at 27 GeV$^2$. The PDFs are reconstructed using our lattice data up to $\avgxxx$ and with the 2-state fits analysis and a 2-parameter fit.}
    \label{fig:xPDF}
\end{figure}

The SU(3) flavor symmetry breaking can be studied by comparing pion and kaon PDFs, which we show in Fig.~\ref{fig:xPDF}. The distributions of the up quarks in both mesons, $x q_\pi^u(x)$ and $x q_K^u(x)$, show a small difference around $x=0.5$ but are otherwise in full agreement for all regions of $x$. This behavior suggests that the up quark is equally prevalent in mesons PDFs and is mostly found in the small- to intermediate-$x$ regions. The strange quark is more prevalent in the kaon than the up quark between $x=0.3$ and $x=0.8$ and has more support in the large-$x$ region. This is what is expected intuitively given the larger strange quark mass. We find the distribution peaks to be $x q_\pi^u(x=0.30)=0.42(5)$, $x q_K^u(x=0.27)=0.43(2)$, and $x q_K^s(x=0.35)=0.52(2)$.

Once the parameters in Table~\ref{table:fit_param_pion_u} have been calculated, the Mellin moments can be extracted from Eq.~(\ref{eq:moments}). This method of calculating $\avgxN$  does not have the problem of operator mixing, which is unavoidable when calculating the moments directly from matrix elements using $n^{\rm th}$-derivative operators for $n>3$. Table~\ref{table:fit_moments} shows the moments for $n\leq6$ calculated from the parameters in Table~\ref{table:fit_param_pion_u} calculated using the 2-parameter fit. The first parenthesis are the statistical and the second are the excited state systematic errors, which are calculated as the difference between the mean moments calculated from the two-state fit and the plateau fit at the largest source-sink separation. We find that our uncertainties are well behaved, even for $n>3$.

\begin{table}[h!]
    \centering
    \begin{tabular}{cccc}
        \hline\hline\\[-2.5ex]
        $q_M^f$ & $\qquad\avgx\qquad$ & $\qquad\avgxx\qquad$   & $\qquad\avgxxx\qquad$  \\[0.5ex]
        \hline\\[-2.5ex]
        $q_\pi^u$  &$\,\,\,$0.229(3)(7)   &$\,\,\,$0.087(5)(7)     &$\,\,\,$0.042(5)(9)    \\
        $q_K^u$  &$\,\,\,$0.217(2)(5)   &$\,\,\,$0.077(2)(1)     &$\,\,\,$0.035(2)(2)    \\
        $q_K^s$  &$\,\,\,$0.279(1)(5)   &$\,\,\,$0.114(2)(4)     &$\,\,\,$0.057(2)(2)    \\[0.5ex]
        \hline\\[-2.5ex]
        $q_M^f$ & $\qquad\avgxxxx\qquad$ & $\qquad\avgxxxxx\qquad$   & $\qquad\langle x^6 \rangle\qquad$  \\[0.5ex]
        \hline\\[-2.5ex]
        $q_\pi^u$  &$\,\,\,$0.023(5)(7)   &$\,\,\,$0.014(4)(6)     &$\,\,\,$0.009(3)(4)  \\
        $q_K^u$  &$\,\,\,$0.019(1)(2)   &$\,\,\,$0.011(1)(1)     &$\,\,\,$0.007(1)(1) \\
        $q_K^s$  &$\,\,\,$0.032(2)(2)   &$\,\,\,$0.020(1)(2)     &$\,\,\,$0.013(1)(2) \\[0.5ex]
        \hline\hline
    \end{tabular}
    \caption{The first six pion and kaon Mellin moments in $\overline{\rm MS}$ at 27 GeV$^2$ as calculated from $\alpha$ and $\beta$ in Table~\ref{tab:params} and Eq.~(\ref{eq:moments}). The first parenthesis contain the statistical error and the second parenthesis contain the systematic error caused by excited-states contamination.}
    \label{table:fit_moments}
\end{table}

\subsection{Comparison with other studies}

\begin{figure}[!h]
    \centering
    \includegraphics[width=\linewidth]{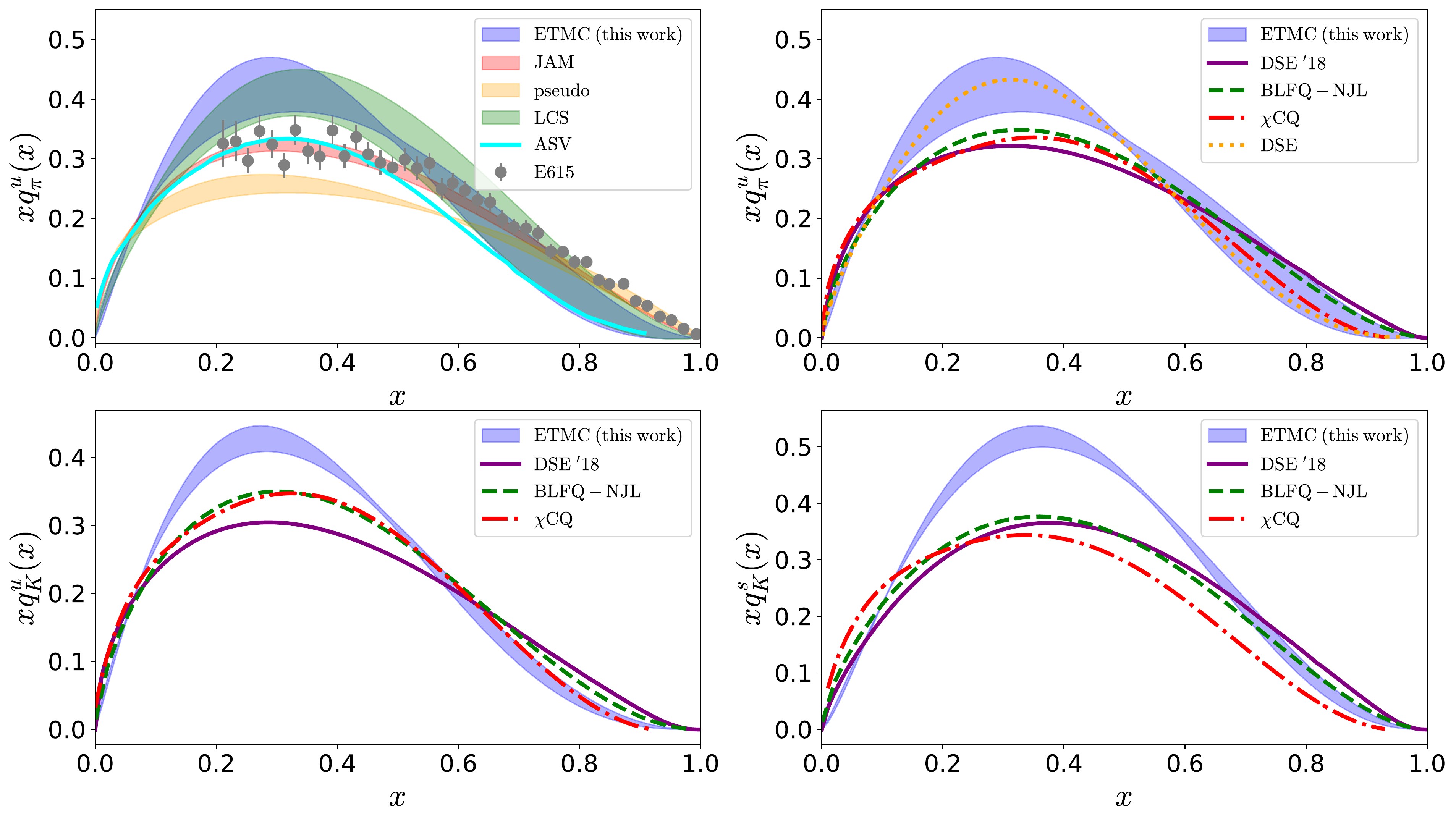}
\caption{Top left panel: Comparison between our results for the pion $x q_\pi^u(x)$ and other studies, all in the $\overline{\rm MS}$ scheme at 27 GeV$^2$. Our results are plotted as the blue band, the E615 data~\cite{Conway:1989fs} as the gray points, the rescaled ASV curve~\cite{Aicher:2010cb} as the cyan line, the JAM global fit as the pink band, the pseudo-ITD~\cite{Joo:2019bzr} lattice results as the orange band, and the current current correlators (LCS) lattice results~\cite{Sufian:2020vzb} as the green band. Top right panel: Comparison between our results for $x q_\pi^u(x)$, DSE~\cite{Chen:2016sno} (orange dotted line), the updated DSE'18~\cite{Bednar:2018mtf} (purple solid line), BLFQ-NJL~\cite{Lan:2019rba} (green dashed line), and $\chi CQ$~\cite{Watanabe:2017pvl} (red dot-dashed line). Bottom panel: Same as top panel for the kaon $x q_K^u(x)$ (left) and $x q_K^s(x)$ (right).}
    \label{fig:PDF_all_fits}
\end{figure}

In Fig.~\ref{fig:PDF_all_fits} we compare our results to some of the other existing calculations of the pion and kaon PDFs. In the left upper panel, our results are compared to lattice results of PDFs calculated with the pseudo-ITD approach~\cite{Joo:2019bzr} and the current-current correlators (LCS) method~\cite{Sufian:2019bol,Sufian:2020vzb}. We find that our results agree better with the LCS method, while the peak from pseudo-ITD study is lower than ours. Our results are also compared to the E615~\cite{Conway:1989fs}, and ASV PDFs~\cite{Aicher:2010cb}, and the JAM Collaboration's global fits~\cite{Barry:2018ort,cao20213dimensional}. In the upper right panel, our results are compared with PDFs calculated using Dyson-Schwinger equations (DSE)~\cite{Chen:2016sno}, the updated DSE '18~\cite{Bednar:2018mtf}, the chiral constituent quark model ($\chi$CQ)~\cite{Watanabe:2017pvl}, the BLFQ Collaboration~\cite{Lan:2019rba} PDFs calculated using light front quantization and QCD evolution (NJL). Calculations of the kaon PDFs are more limited. In the lower panels of Fig.~\ref{fig:PDF_all_fits}, our results for the up quark (left) and strange quark (right) are compared with the $\chi$CQ~\cite{Watanabe:2017pvl}, BLFQ-NJL~\cite{Lan:2019rba}, and DSE '18~\cite{Bednar:2018mtf} results. We find good agreement between all of the studies for the pion and $x q_K^u(x)$ in the small-$x$ region ($x < 0.1$), although the slope of $\chi$CQ at small-$x$ is different from the others. Except for E615, our results are in agreement with the other studies for $x>0.6$. Most of the tension between studies exists in the intermediate region. For the pion, our results are in agreement with DSE~\cite{Chen:2016sno} for all regions of $x$, but we calculate a higher peak than DSE'18. 
In the small- and large-$x$ regions of the kaon, the PDFs agree qualitatively, while in the intermediate-$x$ region, our results are larger than the other results. The kaon comparisons, however, are only qualitative, since no  experimental data exist and non-quantified systematic uncertainties are present in all of the calculations. 

The large-$x$ behavior of the pion and kaon PDFs is an interesting topic of study because different analyses of experimental data and model calculations have found different large-$x$ behavior. For instance, one analysis of the pion Drell-Yan data from the Fermilab E615 experiment~\cite{Conway:1989fs} finds that the PDF falls at large-$x$ like $(1-x)^1$ which corresponds to $\beta=1$, while a more recent analysis~\cite{Aicher:2010cb} suggests the fall to go like $(1-x)^2$ ($\beta=2$). The study using DSE~\cite{Chen:2016sno} also calculates $\beta$ to be closer to 2. Lattice QCD offers a desirable alternative to address this tension. In this work, we find a value for $\beta$ around 2, shown in Table~\ref{table:fit_param_pion_u}. Based on these values, we find that both the pion and kaon PDFs drop off like $\sim(1-x)^2$ at large-$x$.

\section{Summary\label{sec:VII}} 

We use the pion and kaon Mellin moments up to $\avgxxx$ to reconstruct the $x$-dependence of the PDFs. We fit the moments to the standard functional form for PDFs, using 2-parameter and 3-parameter fits and find the 2-parameter fit to be sufficient to reconstruct the PDFs. We study excited-states effects and find non-negligible excited-states contamination for source-sink time separations below 1.5 fm, requiring the use of a two-state fit to control systematic uncertainties. We also test the dependence on the moments included in the fit by reconstructing the PDFs with the highest moment included to be $\avgxx$, $\avgxxx$, or $\avgxxxx$. Our results suggest that the PDF reconstruction is not improved by including $\avgxxxx$. 

We use our PDF results to study pion and kaon structure in multiple ways. The first is the large-$x$ behavior of the PDFs which we find to be $\sim(1-x)^2$. The second is to calculate the higher moments using the fitting parameters and integrals of the functional form of the PDFs. We show results up to $\avgxxxxxx$. Lastly, we study the SU(3) flavor symmetry breaking by comparing the pion PDF to the two quark flavor PDFs of the kaon. We find that the distributions of the light quarks are approximately the same in the pion and kaon and that the strange quark distribution in the kaon is larger in the intermediate- to large-$x$ region. 
 
\acknowledgments

We would like to thank all members of ETMC for a very constructive and enjoyable collaboration. We are grateful to Patrick Barry (JAM Collaboration), Jiangshan Lan and James Vary (BLFQ Collaboration), Raza Suffian (JLAB/W$\&$M), and Akira Watanabe for providing their data for the comparisons presented.
M.C. and C.L. acknowledge financial support by the U.S. Department of Energy Early Career Award under Grant No.\ DE-SC0020405. 
S.B. is supported by the H2020 project PRACE 6-IP (grant agreement No 82376) and the EuroCC project (grant agreement No. 951740).
K.H. is financially supported by the Cyprus Research Promotion foundation under contract number POST-DOC/0718/0100 and the EuroCC project.
This work was in part supported by the U.S. Department of Energy, Office of Science, Office of Nuclear Physics, contract no.~DE-AC02-06CH11357.
This work used computational resources from Extreme Science and Engineering Discovery Environment (XSEDE), which is supported by National Science Foundation grant number TG-PHY170022. 
It also includes calculations carried out on the HPC resources of Temple University, supported in part by the National Science Foundation through major research instrumentation grant number 1625061 and by the US Army Research Laboratory under contract number W911NF-16-2-0189. 

\bibliographystyle{JHEP}
\bibliography{references.bib}

\providecommand{\href}[2]{#2}\begingroup\raggedright\begin{thebibliography}{10}

\bibitem{Hutauruk:2016sug}
P.T.P.~Hutauruk, I.C.~Cloet and A.W.~Thomas, \emph{{Flavor dependence of the
  pion and kaon form factors and parton distribution functions}},
  \href{https://doi.org/10.1103/PhysRevC.94.035201}{\emph{Phys. Rev. C}
  {\bfseries 94} (2016) 035201}
  [\href{https://arxiv.org/abs/1604.02853}{{\ttfamily 1604.02853}}].

\bibitem{Zyla:2020zbs}
{\scshape Particle Data Group} collaboration, \emph{{Review of Particle
  Physics}}, \href{https://doi.org/10.1093/ptep/ptaa104}{\emph{PTEP} {\bfseries
  2020} (2020) 083C01}.

\bibitem{Conway:1989fs}
J.S.~Conway et~al., \emph{{Experimental Study of Muon Pairs Produced by 252-GeV
  Pions on Tungsten}},
  \href{https://doi.org/10.1103/PhysRevD.39.92}{\emph{Phys. Rev.} {\bfseries
  D39} (1989) 92}.

\bibitem{Cichy:2018mum}
K.~Cichy and M.~Constantinou, \emph{{A guide to light-cone PDFs from Lattice
  QCD: an overview of approaches, techniques and results}}, {\emph{accepted in
  Advances of High Energy Physics} (2018) }
  [\href{https://arxiv.org/abs/1811.07248}{{\ttfamily 1811.07248}}].

\bibitem{Constantinou:2020pek}
M.~Constantinou, \emph{{The x-dependence of hadronic parton distributions: A
  review on the progress of lattice QCD}},
  \href{https://doi.org/10.1140/epja/s10050-021-00353-7}{\emph{Eur. Phys. J. A}
  {\bfseries 57} (2021) 77} [\href{https://arxiv.org/abs/2010.02445}{{\ttfamily
  2010.02445}}].

\bibitem{Cichy:2021lih}
K.~Cichy, \emph{{Progress in $x$-dependent partonic distributions from lattice
  QCD}},  in \emph{{38th International Symposium on Lattice Field Theory}}, 10,
  2021 [\href{https://arxiv.org/abs/2110.07440}{{\ttfamily 2110.07440}}].

\bibitem{Detmold:2001dv}
W.~Detmold, W.~Melnitchouk and A.W.~Thomas, \emph{{Parton distributions from
  lattice QCD}}, \href{https://doi.org/10.1007/s1010501c0013}{\emph{Eur. Phys.
  J. direct} {\bfseries 3} (2001) 13}
  [\href{https://arxiv.org/abs/hep-lat/0108002}{{\ttfamily hep-lat/0108002}}].

\bibitem{Holt_2010}
R.J.~Holt and C.D.~Roberts, \emph{Nucleon and pion distribution functions in
  the valence region},
  \href{https://doi.org/10.1103/revmodphys.82.2991}{\emph{Reviews of Modern
  Physics} {\bfseries 82} (2010) 2991–3044}.

\bibitem{Alexandrou:2020gxs}
{\scshape ETM} collaboration, \emph{{Mellin moments $\langle x \rangle$ and
  $\langle x^2 \rangle$ for the pion and kaon from lattice QCD}},
  \href{https://doi.org/10.1103/PhysRevD.103.014508}{\emph{Phys. Rev. D}
  {\bfseries 103} (2021) 014508}
  [\href{https://arxiv.org/abs/2010.03495}{{\ttfamily 2010.03495}}].

\bibitem{Alexandrou:2021}
{\scshape ETM Collaboration} collaboration, \emph{Pion and kaon $⟨{x}^{3}⟩$
  from lattice qcd and pdf reconstruction from mellin moments},
  \href{https://doi.org/10.1103/PhysRevD.104.054504}{\emph{Phys. Rev. D}
  {\bfseries 104} (2021) 054504}
  [\href{https://arxiv.org/abs/2104.02247}{{\ttfamily 2104.02247}}].

\bibitem{Alexandrou:2018egz}
C.~Alexandrou et~al., \emph{{Simulating twisted mass fermions at physical
  light, strange and charm quark masses}},
  \href{https://doi.org/10.1103/PhysRevD.98.054518}{\emph{Phys. Rev.}
  {\bfseries D98} (2018) 054518}
  [\href{https://arxiv.org/abs/1807.00495}{{\ttfamily 1807.00495}}].

\bibitem{Chen:2016sno}
C.~Chen, L.~Chang, C.D.~Roberts, S.~Wan and H.-S.~Zong, \emph{{Valence-quark
  distribution functions in the kaon and pion}},
  \href{https://doi.org/10.1103/PhysRevD.93.074021}{\emph{Phys. Rev. D}
  {\bfseries 93} (2016) 074021}
  [\href{https://arxiv.org/abs/1602.01502}{{\ttfamily 1602.01502}}].

\bibitem{Barry:2018ort}
P.~Barry, N.~Sato, W.~Melnitchouk and C.-R.~Ji, \emph{{First Monte Carlo Global
  QCD Analysis of Pion Parton Distributions}},
  \href{https://doi.org/10.1103/PhysRevLett.121.152001}{\emph{Phys. Rev. Lett.}
  {\bfseries 121} (2018) 152001}
  [\href{https://arxiv.org/abs/1804.01965}{{\ttfamily 1804.01965}}].

\bibitem{Lan:2019rba}
J.~Lan, C.~Mondal, S.~Jia, X.~Zhao and J.P.~Vary, \emph{{Pion and kaon parton
  distribution functions from basis light front quantization and QCD
  evolution}}, \href{https://doi.org/10.1103/PhysRevD.101.034024}{\emph{Phys.
  Rev. D} {\bfseries 101} (2020) 034024}
  [\href{https://arxiv.org/abs/1907.01509}{{\ttfamily 1907.01509}}].

\bibitem{Joo:2019bzr}
B.~Joó, J.~Karpie, K.~Orginos, A.V.~Radyushkin, D.G.~Richards, R.S.~Sufian
  et~al., \emph{{Pion Valence Structure from Ioffe Time Pseudo-Distributions}},
  \href{https://doi.org/10.1103/PhysRevD.100.114512}{\emph{Phys. Rev.}
  {\bfseries D100} (2019) 114512}
  [\href{https://arxiv.org/abs/1909.08517}{{\ttfamily 1909.08517}}].

\bibitem{Sufian:2019bol}
R.S.~Sufian, J.~Karpie, C.~Egerer, K.~Orginos, J.-W.~Qiu and D.G.~Richards,
  \emph{{Pion Valence Quark Distribution from Matrix Element Calculated in
  Lattice QCD}}, \href{https://doi.org/10.1103/PhysRevD.99.074507}{\emph{Phys.
  Rev.} {\bfseries D99} (2019) 074507}
  [\href{https://arxiv.org/abs/1901.03921}{{\ttfamily 1901.03921}}].

\bibitem{Sufian:2020vzb}
R.S.~Sufian, C.~Egerer, J.~Karpie, R.G.~Edwards, B.~Jo\'o, Y.-Q.~Ma et~al.,
  \emph{{Pion Valence Quark Distribution from Current-Current Correlation in
  Lattice QCD}}, \href{https://doi.org/10.1103/PhysRevD.102.054508}{\emph{Phys.
  Rev. D} {\bfseries 102} (2020) 054508}
  [\href{https://arxiv.org/abs/2001.04960}{{\ttfamily 2001.04960}}].

\bibitem{Aicher:2010cb}
M.~Aicher, A.~Schafer and W.~Vogelsang, \emph{{Soft-gluon resummation and the
  valence parton distribution function of the pion}},
  \href{https://doi.org/10.1103/PhysRevLett.105.252003}{\emph{Phys. Rev. Lett.}
  {\bfseries 105} (2010) 252003}
  [\href{https://arxiv.org/abs/1009.2481}{{\ttfamily 1009.2481}}].

\bibitem{cao20213dimensional}
N.Y.~Cao, P.C.~Barry, N.~Sato and W.~Melnitchouk, \emph{Towards the
  3-dimensional parton structure of the pion: integrating transverse momentum
  data into global qcd analysis},  2021.

\bibitem{Bednar:2018mtf}
K.D.~Bednar, I.C.~Clo\"et and P.C.~Tandy, \emph{{Distinguishing Quarks and
  Gluons in Pion and Kaon Parton Distribution Functions}},
  \href{https://doi.org/10.1103/PhysRevLett.124.042002}{\emph{Phys. Rev. Lett.}
  {\bfseries 124} (2020) 042002}
  [\href{https://arxiv.org/abs/1811.12310}{{\ttfamily 1811.12310}}].

\bibitem{Watanabe:2017pvl}
A.~Watanabe, T.~Sawada and C.W.~Kao, \emph{{Kaon quark distribution functions
  in the chiral constituent quark model}},
  \href{https://doi.org/10.1103/PhysRevD.97.074015}{\emph{Phys. Rev. D}
  {\bfseries 97} (2018) 074015}
  [\href{https://arxiv.org/abs/1710.09529}{{\ttfamily 1710.09529}}].

\end{thebibliography}\endgroup

\end{document}